\documentclass[aps,preprintnumbers,amsmath,amssymb,twocolumn]{revtex4-1}

\usepackage{graphicx}
\usepackage{dcolumn}
\usepackage{bm}
\usepackage{color}
\usepackage{amssymb}   
\usepackage{amsthm}

\begin{document}

\title{Dynamical density functional theory analysis of the laning instability in sheared soft matter}

\author{A. Scacchi$^1$}
\author{A.J. Archer$^2$}
\author{J.M. Brader$^1$}

\affiliation{1. Department of Physics, University of Fribourg, 1700 Fribourg, Switzerland}
\affiliation{2. Department of Mathematical Sciences, Loughborough University, Loughborough LE11 3TU, UK}

\begin{abstract}
Using dynamical density functional theory (DDFT) methods we investigate the laning instability of a sheared colloidal suspension. The nonequilibrium ordering at the laning transition is driven by non-affine particle motion arising from interparticle interactions. Starting from a DDFT which incorporates the non-affine motion, we perform a linear stability analysis that enables identification of the regions of parameter space where lanes form. We illustrate our general approach by applying it to a simple one-component fluid of soft penetrable particles. 
\end{abstract}
\maketitle

\section{Introduction}

When colloidal suspensions are driven by a shear-flow, if the shearing is strong enough, the particles can exhibit a laning transition that is sometimes described as the particles forming ``strings'' or ``layered'' structures. The colloidal particles self-organise into these structures in order to slide past one another more efficiently. These have been observed both for oscillatory and steady shear \citep{ackerson_pusey_1988, corte2008random, besseling, nikoubashman2011cluster, likos_nucleation, pore_study}. In the work of Besseling et al.\ \citep{besseling} the sliding layer state is identified as being a distinct phase from the string phase and they also observe a tilted layer phase. In \citep{besseling} the system studied was suspensions of hard-sphere-like PMMA colloids. The structures formed under shear were compared with results from Brownian Dynamics (BD) computer simulations for hard spheres and good agreement was observed. The formation of lanes under shear has not just been observed for hard-sphere systems: for example in \citep{pore_study} a dense system of charged colloids modelled as interacting via a Yukawa pair potential was studied using BD and when exposed to shear flow in a confined (slit-pore) geometry, laned states were observed and in \citep{nikoubashman2011cluster, likos_nucleation} strings parallel to the flow direction form in a system of soft penetrable particles. Note that BD simulations do not include the hydrodynamic interactions between the particles mediated via the background solvent. However, one should in general expect in some systems or in other regimes that the hydrodynamic interactions could be important and might suppress the laning \citep{foss_brady_2000}. Clearly, laning transitions can significantly affect the rheology.

There is a closely related laning instability observed in binary systems of colloidal particles that are driven in opposite directions \citep{lowen2010particle}. This can for example occur because the buoyant mass of the two species has the opposite sign or because the colloids are oppositely charged and they are driven by an electric field. This effect was observed both in experiments \citep{leunissen2005ionic} and analysed theoretically \cite{chakrabarti2003dynamical, lane_mixture}.

An early theoretical analysis to understand the density distribution of the colloids in the laned state was made by Hoffman \cite{hoffman1974discontinuous}. A natural theoretical framework for determining the density distribution is dynamical density functional theory (DDFT) \cite{marconi_tarazona_1, marconi_tarazona_2, archer_evans, ArRa04}, a theory for the collective dynamics of Brownian particles. As originally formulated, the theory assumes the background solvent is stationary, but it is straightforward to include a background flow by the addition of an advection term \cite{rauscher_penna_2007}. However, this theory is still unable to describe the laning transition. Using symmetry and other arguments, an extra contribution to the particle currents in the DDFT was argued for in \citep{chakrabarti2003dynamical} for the case of binary mixtures of oppositely driven particles. This contribution captures the coupling between flow and interparticle interactions that generates lateral fluxes perpendicular to the flow, leading to a DDFT able to describe the laning transition.

A more general theory for these effects \citep{brader_kruger_2011} can be obtained by considering the action of one particle moving past another, described by a flow kernel. The authors of Ref.~\citep{brader_kruger_2011} applied the theory to describe hard spheres under simple shear, showing that under strong enough driving the particles order into layers parallel to the confining walls. The transition to the laned state is manifest as the onset of an instability in the bulk uniform density state to form a periodically modulated density distribution. More recently, using a flow-kernel that is exact in the low-density limit, this theory was applied to study shear-induced migration and laning of hard spheres under Poiseuille flow \citep{scacchi_kruger_brader_2016}.

Here, we perform a linear stability analysis of the advected DDFT that incorporates a flow kernel, to obtain a general expression for the dispersion relation in sheared colloidal suspensions. A similar linear stability analysis has ben performed previously for the original DDFT \cite{archer_evans, evans_79} and also applied to study solidification fronts and to calculate front speeds via a marginal stability analysis \cite{archer2012solidification, andy_walters_thiele_knobloch}. The linearisation is formally exact and allows to determine the onset threshold for the laning instability. However, it requires as input the flow kernel, which is not known exactly. Nonetheless, the derivation gives good insight into the origin of the laning instability and shows that the laning transition corresponds to the linear-instability threshold of the uniform density fluid state.

We also apply the approach to study the laning transition in sheared fluids of soft core particles. The particular model system we study is the so-called generalized exponential models of index $n$ (GEM-$n$) model \citep{likos2001effective, mladek2006formation, moreno, andy_walters_thiele_knobloch, andy_alexandr}, with index $n=8$, which is a simple model for dendrimers in solution \citep{likos2001effective, lenz2012microscopically}. Using dimensional analysis and physical arguments we propose a simple approximation for the shear-kernel for such soft-core systems. For hard spheres one can consider the dynamics of one sphere rolling over the other to obtain approximations for the shear-kernel \citep{brader_kruger_2011, scacchi_kruger_brader_2016}, but such arguments are not applicable to soft particles. Our theory predicts a laning transition and using a simple mean-field approximation for the Helmholtz free energy functional, we calculate the liquid density profiles for the case where the fluid is sheared between two planar walls. These results show that there is a transition in the density profiles to a state with periodic density oscillations that have a finite amplitude at all distances from the walls and this laning transition occurs precisely where it is predicted by the linear stability analysis.

This paper is structured as follows: In Sec.\ \ref{sec:2} we introduce the DDFT for sheared suspensions. In Sec.\ \ref{sec:3} we perform the linear stability analysis to obtain the dispersion relation for a sheared suspension. In Sec.\ \ref{sec:4} we introduce the GEM-$n$ model fluid and postulate a suitable approximation for the flow kernel for soft particles, illustrating the resulting GEM-8 flow kernel in both 2 and 3 dimensions. In Sec.\ \ref{sec:5} we present density profiles for the GEM-8 fluid sheared between two parallel planar walls and plot the dispersion relation and the phase diagram displaying the location of the laning transition for various shear rates. Finally, in Sec.\ \ref{sec:6}, we discuss our results and draw our conclusions.

\section{DDFT for sheared suspension}
\label{sec:2}

Dynamical density functional theory is a theory for overdamped Brownian particles suspended in a fluid and can be derived 
either from the Langevin equation of motion~\cite{marconi_tarazona_1, marconi_tarazona_2} or from the Smoluchowski equation \citep{archer_evans, ArRa04}. When the surrounding fluid has a non-zero velocity, then an advected form of DDFT can be derived~\citep{rauscher_penna_2007}. This continuity type of equation has the form:

\begin{equation}
\frac{\partial \rho(\textbf{r},t)}{\partial t}+\nabla\cdot\left[\rho(\textbf{r},t) \textbf{v}(\textbf{r})\right]=\Gamma\nabla\cdot\left[\rho(\textbf{r},t)\nabla\frac{\delta \mathcal{F[\rho]}}{\delta\rho(\textbf{r},t)}\right],\label{DDFT}
\end{equation}
where $\Gamma$ is the mobility, $\textbf{v}$ is the solvent velocity, and $\mathcal{F}$ the equilibrium Helmholtz free energy functional 
\cite{evans_79, evans_92, hansen_mcdonald}. The Helmholtz free energy is composed of three terms:
\begin{align}
  \mathcal{F}[\rho(\textbf{r})]=\mathcal{F}^{id}[\rho(\textbf{r})] 
+ \mathcal{F}^{exc}[\rho(\textbf{r})] +
  \int d\textbf{r}\rho(\textbf{r})V_{ext}(\textbf{r}),
\label{helmholtz}  
\end{align}
where $V_{ext}$ is the external potential and where the ideal gas part is given exactly by 
\begin{equation}\label{ideal}
\mathcal{F}^{id}[\rho(\textbf{r})]=\int d\textbf{r} \rho(\textbf{r})(\ln[\Lambda^3\rho(\textbf{r})]-1),
\end{equation}
where $\Lambda$ is the thermal wavelength. $\mathcal{F}^{exc}[\rho]$ is the excess free energy, the contribution due to the interactions between the particles. Thus, the functional derivative in Eq.~(\ref{DDFT}) is
\begin{equation}
\frac{\delta\mathcal{F}[\rho(\textbf{r})]}{\delta\rho(\textbf{r})}=k_BT\ln[\Lambda^3\rho(\textbf{r})]-k_BTc^{(1)}(\textbf{r})+V_{ext}(\textbf{r}),
\end{equation}
where
\begin{equation}
c^{(1)}(\textbf{r})\equiv -\beta \frac{\delta\mathcal{F}^{exc}[\rho(\textbf{r})]}{\delta\rho(\textbf{r})}
\end{equation}
is the one-body direct correlation function~\citep{evans_79,evans_92, hansen_mcdonald}. 
We consider the case analysed in~\citep{brader_kruger_2011} of a fluid sheared between to parallel planar walls. We assume without loss of generality that the normal vector to the walls points in the direction parallel to the $y$-axis. If we assume that the external flow field is a simple steady shear with flow in the $x$-direction and a linear gradient in the $y$-direction, i.e. $\textbf{v}(\textbf{r})=\textbf{v}^{aff}(\textbf{r})=y\dot{\gamma}\textbf{e}_x$, where $\dot{\gamma}$ is the shear rate, then these two properties together lead to a vanishing advection term in Eq. (\ref{DDFT}), i.e.\ $\nabla\cdot\left[\rho(\textbf{r},t) \textbf{v}(\textbf{r})\right]=0$, resulting in a unchanged density profile under shear. This is well-known not to always be the case, so there must be an additional contribution to $\textbf{v}(\textbf{r})$. The solution to this, proposed in \citep{brader_kruger_2011}, is to write the velocity field as the sum of the affine flow and a particle induced fluctuation flow so that $\textbf{v}(\textbf{r})=\textbf{v}^{aff}(\textbf{r})+\textbf{v}^{fl}(\textbf{r})$. The term $\textbf{v}^{fl}(\textbf{r})$ ensures that a pair of approaching particles experiencing different velocities feel the correct force by 'flowing around€™ each other. The quantity $\textbf{v}^{fl}(\textbf{r})$ is not known exactly, however it must be a functional of the fluid density and it must also be zero when the fluid density is uniform, $\rho(\textbf{r})=\rho_b$. Assuming we can make a functional Taylor expansion, we write
\begin{equation}
\textbf{v}^{fl}[\rho(\textbf{r})]=\textbf{v}^{fl}[\rho_b]+\int d\textbf{r}' \delta\rho(\textbf{r}') \frac{\delta \textbf{v}^{fl}[\rho(\textbf{r})]}{\delta\rho(\textbf{r}')}\bigg|_{\rho_b}+\mathcal{O}(\delta\rho^2),
\end{equation}
where $\delta\rho(\textbf{r})=\rho(\textbf{r})-\rho_b$. The first term is zero and truncating after the second term we obtain the form suggested in Ref.~\cite{brader_kruger_2011}, i.e.
\begin{equation}
\textbf{v}^{fl}(\textbf{r})=\int d\textbf{r}'\delta\rho(\textbf{r}')\boldsymbol{\kappa}(\textbf{r}-\textbf{r}'),\label{non-affine}
\end{equation}
where $\boldsymbol{\kappa}(\textbf{r}-\textbf{r}')=\frac{\delta \textbf{v}^{fl}[\rho(\textbf{r})]}{\delta\rho(\textbf{r}')}\big|_{\rho_b}$ is the flow kernel. This kernel function describes the velocity of a particle colliding with a neighbour. In the bulk, the kernel function is translationally invariant, but in the presence of confinement, this is no longer the case, especially when the distance between the particle and the confining substrate is on the order of the particles diameter. Physical interpretation of this function was given in~\cite{brader_kruger_2011}, where the authors derived it in the case of hard spheres, by a geometrical construction. More recently in~\cite{scacchi_kruger_brader_2016} the authors provide a derivation of the kernel function from an exactly solvable low density limit. Both of these works apply to hard-spheres. Here our interest is more general and later in this paper we postulate a physically motivated expression [equation \eqref{kernel_grad_phi}] for the flow kernel that is applicable in the case of soft particles.

\section{Stability analysis}\label{stability_analysis} 
\label{sec:3}

The following derivation is a general linear stability analysis for the generic DDFT with non-affine advection. We start by writing the one-body density as a bulk density plus a small perturbation as follows:
\begin{equation}
\rho(\textbf{r},t)=\rho_b+\delta\rho(\textbf{r},t)=\rho_b+\sum_{\textbf{k}}\epsilon_{\textbf{k}} e^{i\textbf{k}\cdot \textbf{r}+\omega t},\label{density_perturbation}
\end{equation}
where $\textbf{k}$ is a wave vector, $\epsilon_{\textbf{k}}$ a corresponding amplitude and $\omega(\textbf{k})$ a growth/decay rate, referred to as the dispersion relation. Without loss of generality, we now consider the stability perpendicular to the flow direction and reduce the problem to a 2-dimensional (2D) system in the $(x,y)$ plane. It is convenient to write the velocity field as:
\begin{equation}\label{velocity}
\textbf{v}(\textbf{r})=v^{aff}(y)\textbf{e}_{x}+\textbf{v}^{fl}(\textbf{r})=\dot{\gamma} y \textbf{e}_x + \textbf{v}^{fl}(\textbf{r}),
\end{equation}
where we have assumed an affine simple shear acting along the $x$-axis, where $\dot{\gamma}$ is the shear rate and $\textbf{v}^{fl}(\textbf{r})$ is the non-affine term. Using these last definitions, and a functional Taylor expansion of $c^{(1)}(\textbf{r})$, i.e.
\begin{equation}
c^{(1)}(\textbf{r})=c^{(1)}[\rho_b]+\int d\textbf{r}' \delta\rho(\textbf{r}')c^{(2)}(\mid \textbf{r}-\textbf{r}'\mid)+\mathcal{O}(\delta\rho^2),
\end{equation}
where $c^{(2)}(r)=\delta c^{(1)}/\delta\rho\mid_{\rho_b}$ is the bulk fluid two-body direct correlation function, Eq. (\ref{DDFT}) reduces to:
\begin{equation}
\begin{split}
\omega\delta\rho(\textbf{r},t)&+\rho_b\nabla\cdot\textbf{v}(\textbf{r})+\nabla\cdot\left[\delta\rho(\textbf{r},t)\textbf{v}(\textbf{r})\right]\\&=-k^2 D\left[1-\rho_b\hat{c}(k)\right]\delta\rho(\textbf{r},t)+\mathcal{O}(\delta \rho^2),
\end{split}
\end{equation}
where $D=k_B T\Gamma$ is the diffusion coefficient and $\hat{c}(k)$ is the Fourier transform of $c^{(2)}(r)$. The right hand side of the last equation is a known result~\cite{marconi_tarazona_1, archer_evans, archer2012solidification, evans_79, andy_walters_thiele_knobloch}. The Taylor expansion of the free energy is an appropriate approximation, since we assume the density variations around the bulk value to be small, i.e. $\delta\rho\ll1$. We then recall that for an equilibrium fluid $\left[1-\rho_b\hat{c}(k)\right]=1/S(k)$, where $S(k)$ is the static structure factor~\cite{evans_92}. 
Using Eq. (\ref{velocity}) we obtain:
\begin{equation}
\begin{split}
\omega\delta\rho(\textbf{r},t)+\rho_b\nabla\cdot\textbf{v}^{fl}(\textbf{r})&+i\textbf{k}\cdot\textbf{v}\delta\rho(\textbf{r},t)+\delta\rho(\textbf{r},t)\nabla\cdot\textbf{v}^{fl}(\textbf{r})\\&=-\frac{k^2D}{S(k)}\delta\rho(\textbf{r},t)+\mathcal{O}(\delta \rho^2).
\end{split}
\end{equation}
Dividing each term by $\delta\rho(\textbf{r},t)$ we get the following expression for $\omega$:
\begin{equation}
\begin{split}
\omega&=-\frac{\rho_b}{\delta\rho(\textbf{r},t)}\nabla\cdot\textbf{v}^{fl}(\textbf{r})-ik_x v^{aff}_x(\textbf{r})-ik_x v^{fl}_x(\textbf{r})\\&-ik_y v^{fl}_y(\textbf{r})-\nabla\cdot\textbf{v}^{fl}(\textbf{r})-\frac{k^2D}{S(k)}+\mathcal{O}(\delta \rho).
\end{split}
\end{equation}
We are interested in the growth/decay of perturbations perpendicular to the flow, so we consider wave vectors $\textbf{k}=(0,k_y)$. The last equation now takes the easier form:
\begin{equation}
\begin{split}
\omega&=-\frac{\rho_b}{\delta\rho(\textbf{r},t)}\nabla\cdot\textbf{v}^{fl}(\textbf{r})-ik_y v^{fl}_y(\textbf{r})-\nabla\cdot\textbf{v}^{fl}(\textbf{r})\\&-\frac{k_y^2D}{S(k_y)}+\mathcal{O}(\delta \rho).
\label{omega1}
\end{split}
\end{equation}
The non-affine velocity term has the general convolution form in Eq.\ (\ref{non-affine}). Using Eq.\ (\ref{density_perturbation}) in Eq.\ (\ref{non-affine}), and using $\textbf{r}''=\textbf{r}-\textbf{r}'$ we get:
\begin{equation}
\begin{split}
\textbf{v}^{fl}(\textbf{r})&=\int d\textbf{r}''\delta \rho(\textbf{r}-\textbf{r}'',t)\boldsymbol{\kappa}(\textbf{r}'')\\&=\sum_{\textbf{k}}\epsilon_{\textbf{k}} e^{\omega t} \int d\textbf{r}''e^{i\textbf{k}\cdot(\textbf{r}-\textbf{r}'')}\boldsymbol{\kappa}(\textbf{r}'')\\&=\delta\rho(\textbf{r},t)\int d\textbf{r}''e^{-i\textbf{k}\cdot\textbf{r}''}\boldsymbol{\kappa}(\textbf{r}'')\\&=\delta\rho(\textbf{r},t)\int d\textbf{r}''\left[\cos(\textbf{k}\cdot\textbf{r}'')-i\sin(\textbf{k}\cdot\textbf{r}'')\right]\boldsymbol{\kappa}(\textbf{r}'').
 \end{split}
\end{equation}
We recall that the $y$-component of the kernel function, $\kappa_y$, is an odd function, which implies that $\int d\textbf{r}' \kappa_y(\textbf{r}-\textbf{r}')=0$ and also that the first of the above integrals is vanishing, so that we can simplify to obtain:
\begin{equation}
\begin{split}
\textbf{v}^{fl}(\textbf{r})&=-i\delta\rho(\textbf{r},t)\int d\textbf{r}'' \sin(\textbf{k}\cdot\textbf{r}'')\boldsymbol{\kappa}(\textbf{r}'')\\&=-i\delta\rho(\textbf{r},t)\boldsymbol{\alpha}(\textbf{k}),\label{v_fl_integral}
\end{split}
\end{equation}
where $\boldsymbol{\alpha}(\textbf{k})=\int d\textbf{r}\sin(\textbf{k}\cdot\textbf{r})\boldsymbol{\kappa}(\textbf{r})$. In equation (\ref{omega1}), the second and the third term in the right hand side are of order $\delta\rho$, which can be neglected, since we consider the limit of small perturbation $\delta\rho$, i.e.\ $\epsilon_{\textbf{k}}\rightarrow 0$. We are left with the equation:
\begin{equation}
\begin{split}
\omega&=-\frac{\rho_b}{\delta\rho(\textbf{r},t)}\nabla\cdot\textbf{v}^{fl}-\frac{k_y^2D}{S(k_y)}\\&=i\frac{\rho_b}{\delta\rho(\textbf{r},t)}\nabla\cdot\left[\delta\rho(\textbf{r},t) \boldsymbol{\alpha}(\textbf{k})\right]-\frac{k_y^2D}{S(k_y)}\\&=
i\boldsymbol{\alpha}(\textbf{k})\frac{\rho_b}{\delta\rho(\textbf{r},t)}\nabla\delta\rho(\textbf{r},t)-\frac{k_y^2D}{S(k_y)}\\&=-\rho_b \textbf{k}\cdot\boldsymbol{\alpha}(\textbf{k})-\frac{k_y^2D}{S(k_y)}.
\end{split}
\end{equation} 
Since we have chosen to consider the wave vectors $\textbf{k}=(0,k_y)$ in order to study the instability along the $y$-axis, we can write:
\begin{equation}
\omega(0, k_y)=-\rho_b k_y \alpha_y(k_y)-\frac{k_y^2D}{S(k_y)},
\label{dispersion_relation_final}
\end{equation}
where $\alpha_y(k_y)=\int d\textbf{r} \sin(k_y y)\kappa_y(\textbf{r})$. When $\dot{\gamma}=0$, then $ \alpha_y(k_y)=0$ and so since the static structure factor $S(k)>0$ for all $k$, Eq.\ (\ref{dispersion_relation_final}) shows that $\omega<0$ for all $k$, which, of course, means that when $\dot{\gamma}=0$ the uniform liquid is linearly stable. However, for $\dot{\gamma}>0$ this is not necessarily the case. Recall that the static structure factor $S(k_y)$ typically has a peak at $k_y\sim 2\pi/\sigma$, where $\sigma$ is the typical distance between pairs of neighbouring particles~\citep{hansen_mcdonald}. This implies there is a peak at $k_y\sim 2\pi/\sigma$ in the second term on the right hand side of Eq.~(\ref{dispersion_relation_final}). For $\dot{\gamma}>0$ the effect of the function $\alpha_y(k_y)$, and thus of the function $\kappa_y$, is to increase the height of this peak. For increasing shear rate $\dot{\gamma}$ the height of the peak in $\omega(0,k_y)$ must increase till eventually we have $\omega(0, k_y\sim 2\pi/\sigma)=0$. This means that the uniform density state is marginally unstable. For even greater shear rate $\dot{\gamma}$ we have $\omega(0, k_y\sim 2\pi/\sigma)>0$ and so the uniform liquid becomes linearly unstable. This corresponds to the onset of laning. For an illustration of this general observation for a particular model fluid see below (in particular, in Fig.~\ref{fig_dispersion}). While the above linear stability analysis is able to predict the threshold for the onset of the laning transition, it can not predict the amplitude of the density oscillations in the laned state. To obtain the amplitude, solution of the full nonlinear theory is required. We now illustrate these general observations for a particular model fluid.


\section{The GEM-$n$ fluid}
\label{sec:4}

The GEM-$n$ is a simple model for polymeric macromolecules in a solvent. The effective interaction potential between the centres of mass of the polymers is modelled by a repulsive potential with the form \citep{likos2001effective, mladek2006formation, moreno, andy_walters_thiele_knobloch, andy_alexandr, lenz2012microscopically}
\begin{equation}
\phi(r)=\epsilon \exp[-(r/R)^n],
\end{equation}
where the parameter $\epsilon>0$ determines the strength of the repulsion between pairs of particles, the index $n$ determines how rapidly the potential decays outside of the core, which has radius $R$. This is roughly the polymer radius of gyration. Since this potential is purely repulsive, the fluid does not exhibit liquid-gas phase separation. However, the GEM-$n$ does exhibit a fluid to crystal phase transition. The important difference between a GEM-$n$ system and system where the particles have a hard core such as hard-spheres, is that the particles can overlap, giving rise to cluster crystals \citep{mladek2006formation}, due to the particle penetrability. This system is relatively well understood, and aspects such as how the dynamics of solidification proceeds \citep{andy_walters_thiele_knobloch}, behaviour under shear \citep{nikoubashman2011cluster, likos_nucleation} and crystallisation under confinement \citep{andy_alexandr, moreno} have been addressed. Another important aspect of this model is that a simple mean-field approximation for the Helmholtz free energy is surprisingly accurate, i.e.\ we can use
\begin{align}
  \mathcal{F}^{exc}[\rho]= \frac{1}{2}\int d\textbf{r} \int d\textbf{r}' \rho(\textbf{r})\rho(\textbf{r}')\phi(\mid\textbf{r}-\textbf{r}'\mid). 
\label{meanfield}  
\end{align}

We turn now to the physics of the flow kernel $\boldsymbol{\kappa}(\textbf{r})$ for such soft particles. As discussed in Refs.\ \citep{brader_kruger_2011, scacchi_kruger_brader_2016} the flow kernel describes the force two particles exert on one another as they move past one another due to a shear flow. Therefore, for soft particles a natural assumption is that $\boldsymbol{\kappa}(\textbf{r})\propto-\nabla\phi(\textbf{r})$, where $\phi$ is the pair interaction potential between the particles. Since $\boldsymbol{\kappa}(\textbf{r})=0$ when $\dot{\gamma}=0$, but is non-zero otherwise, it is also natural to assume that $\boldsymbol{\kappa}(\textbf{r})$ is proportional to the shear rate $\dot{\gamma}$. From Eq.\ (\ref{non-affine}) we see that $\boldsymbol{\kappa}(\textbf{r})$ has the dimensions of a velocity, so there must be a pre-factor, which on dimensional grounds we assume to be of the form $f(\rho_b)R^2\beta$, where $f(\rho_b)$ is a dimensionless quantity, that in principle can depend on the fluid density. Combining all these considerations we obtain the following pair-force weighted with the shear rate approximation for the shear kernel
\begin{equation}
\boldsymbol{\kappa}(\textbf{r})=-f(\rho_b)R^2\dot{\gamma}\nabla \beta\phi(\textbf{r}).\label{kernel_grad_phi}
\end{equation}
One approach could be to treat the dimensionless function $f(\rho_b)$ as a fitting function to match results from a different method, such as from BD computer simulations. However, since at this stage we can not discern anymore about this function, we assume $f=1$.

\subsection{Flow kernel for uniaxial shear}\label{subsec:4.A}

\begin{figure}[t!]
\includegraphics[scale=.47]{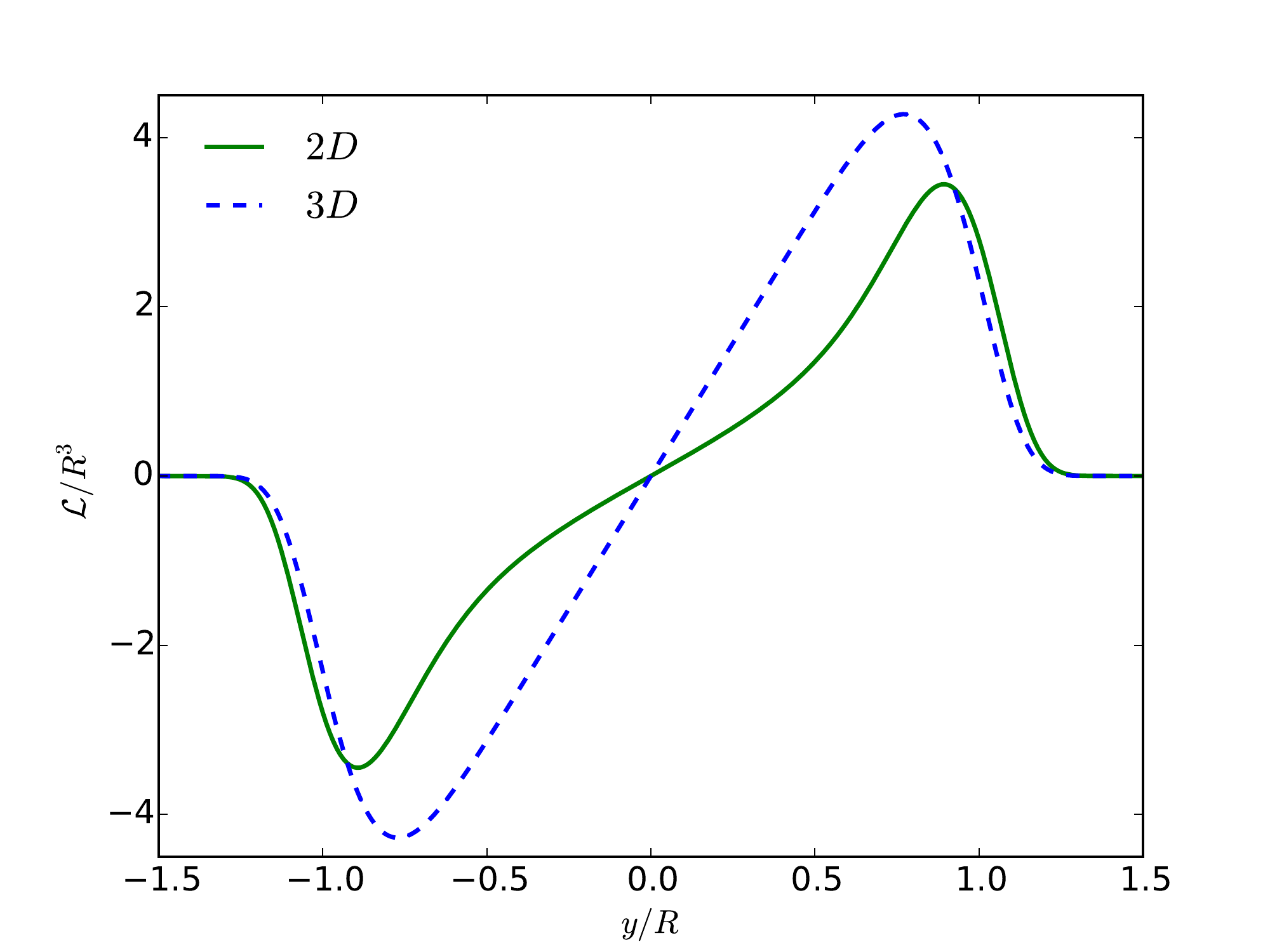}\caption{The reduced flow kernel $\mathcal{L}$ for both the 2D and 3D GEM-8 fluid, for $\beta\epsilon=1$.}\label{kernels}
\end{figure}

When the shear is uniaxial, i.e.\ when $\textbf{v}^{aff}(\textbf{r})=y\dot{\gamma}\textbf{e}_x$, we can reduce the situation to an effective 1-dimensional (1D) situation, by assuming that the density varies only along the stability axis, i.e.\ perpendicular to the walls responsible for the shear. For the GEM-8 fluid, the details of evaluating the integrals in Eq.\ (\ref{v_fl_integral}) with the flow kernel given in (\ref{kernel_grad_phi}) can be found in the Appendix for both the case when the fluid is 3-dimensional (3D) and when the fluid is 2D. That some of the integrals do not depend on $\rho(\textbf{r})=\rho(y)$ means is that the velocity fluctuation term can be written as:
\begin{equation}
v^{fl}_y(y)=\dot{\gamma}\int_{-\infty}^{\infty} \rho(y-y') \mathcal{L}(y') dy'.
\end{equation}
The function $\mathcal{L}(y)$ can be numerically calculated in advance of the main calculation, reducing the full 2D or 3D problem to an effective 1D one, that only varies in $y$. In Fig.~\ref{kernels} we compare the reduced flow kernel $\mathcal{L}(y)$ for the GEM-8 fluid in 2D with that in 3D. We see that in both cases $\mathcal{L}(y)$ is an odd function, which is a general property of the flow kernel \citep{brader_kruger_2011, scacchi_kruger_brader_2016}.

\section{Results}
\label{sec:5}


\subsection{Sheared 2D GEM-8 fluid}\label{subsec:5.A}

\begin{figure}[t!]
\centering
\includegraphics[scale=.40]{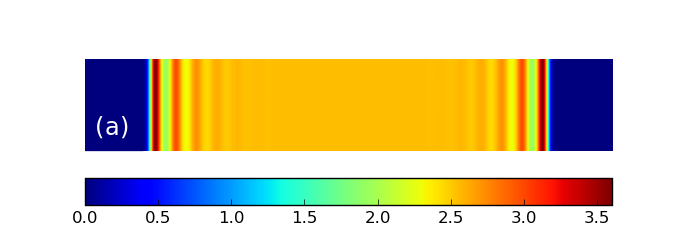}
\includegraphics[scale=.40]{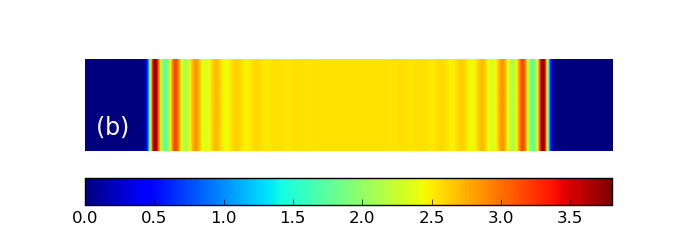}
\caption{(a) 2D equilibrium fluid density distribution for the GEM-8 system. (b) the non-equilibrium sheared system, with shear rate $\dot{\gamma}=0.146$. In both, the bulk density $\rho_b R^2=2.55$ and $\beta\epsilon=1$. The distance between the walls is $L_y=28.5R$ and the vertical system size is $5.42R$.}\label{2D_eq}
\end{figure}

In this section we present results for the 2D GEM-8 system. We assume that the confining walls have smooth surfaces, i.e.\ the potentials only vary in the $y$-direction, so that $V_{ext}(\textbf{r})=V_{ext}(y)$, which is a requirement for the assumptions made in Sec.\ \ref{sec:3} to be true. We choose the following external potential:
\begin{equation}
V_{ext}(y)=
\begin{cases}
V_0 e^{-\frac{y^2}{2R_w^2}}, &\mbox{$y>0$ and $y<L_y/2$} 
\\ 
\infty , &\mbox{$y<0$ or $y>L_y$}
\\
V_0 e^{-\frac{(y-L_y)^2}{2R_w^2}}, &\mbox{$y>L_y/2$ and $y<L_y$}

\end{cases},\label{external_potential}
\end{equation}
where $L_y$ is the distance between the walls, $R_w$ is the range of the wall potentials and $V_0$ is the repulsion strength. This potential corresponds to a pair of parallel soft purely repulsive walls. The softness aids the numerical stability of the calculations. In Fig.~\ref{2D_eq} we present results for a fluid confined between walls with $L_y=28.5R$, $\beta V_0=120$ and $R_w=0.9R$ and with temperature $k_BT/\epsilon=1$ and bulk density $\rho_b R^2=2.55$ (i.e.\ this is the density in the middle of the slit). Along the $x$-axis we impose periodic boundary conditions and set the vertical system size to be $5.42R$. In Fig.~\ref{2D_eq}(a) we show the equilibrium fluid density profile. We observe oscillations in the fluid density at the walls, due to packing effects. The amplitude of the oscillations decays relatively quickly as one moves towards the centre of the system, where the density takes the bulk value.

In Fig.~\ref{2D_eq}(b) we display the non-equilibrium density profile when the fluid is sheared, with shear rate $\dot{\gamma}>0$. Henceforth we give the shear rate in terms of the dimensionless shear rate $\dot{\gamma}^*=\dot{\gamma}R^2/D$. We also henceforth drop the superscript `*'. The results in Fig.~\ref{2D_eq}(b) are for $\dot{\gamma}=0.146$. We see that the effect of the shear is to make the amplitude of the density oscillations greater; i.e.\ the peaks of the density maxima are higher than in the case when $\dot{\gamma}=0$ and the troughs are lower. Also, the locations of the peaks are slightly shifted towards the walls. This is due to the softness of the interfaces. Note that in this $\dot{\gamma}>0$ case the density profile remains invariant in the vertical $x$-direction.

\begin{figure}[t!]
\includegraphics[scale=.47]{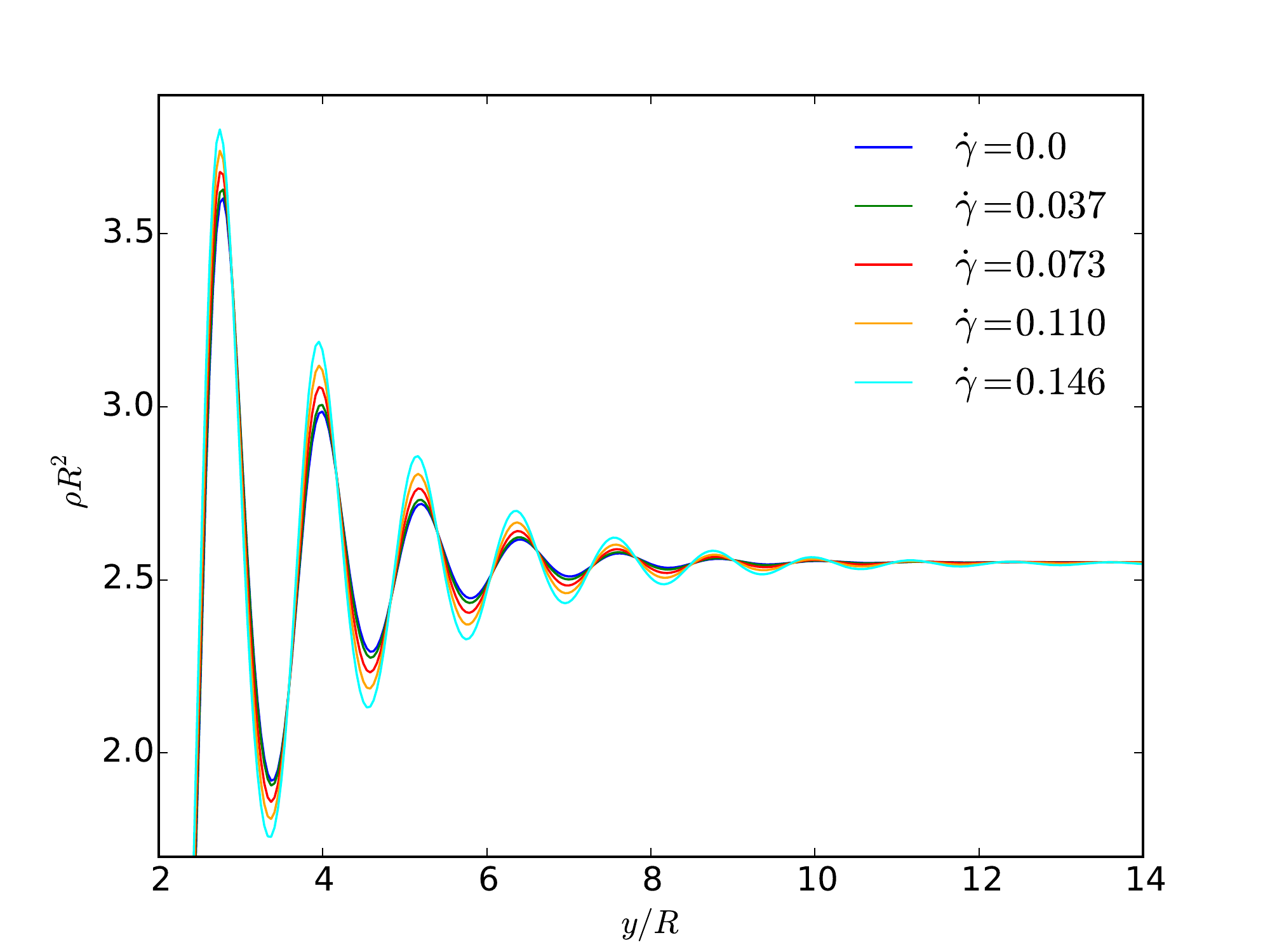}\caption{Density profiles for a 2D GEM-8 system under shear with bulk density $\rho_b R^2=2.55$ and $\beta\epsilon=1$. The distance between the walls is $L_y=28.5R$ and the vertical system size is $5.42R$. The amplitude of the density oscillations at the walls grow monotonically as a function of the shear rate $\dot{\gamma}$, and the peaks are slightly shifted towards the wall.}\label{2D_1D}
\end{figure}

Since the profiles are invariant in the $x$-direction, in Fig.~\ref{2D_1D} we display a cut through the profiles at the wall along the $y$-direction for different shear rates $\dot{\gamma}$, which allows for a better understanding of the influence of $\dot{\gamma}$ on the structure of the liquid.

These 2D calculations are computationally demanding, and give the same information that can be obtained from assuming the density only varies in the $y$-direction. This justifies the assumptions made in Sec.\ \ref{stability_analysis} and that at least in some cases that one can assume $\rho(\textbf{r})=\rho(y)$. We do not expect this to always be true, but for simple shear we expect that this will usually be the case, especially near the onset of laning.

\subsection{Dispersion relation and laning transition}\label{subsec:5.B}

The dispersion relation $\omega$ in Eq.\ (\ref{dispersion_relation_final}) contains all the information required to determine if the uniform liquid is stable or not under a given external shear rate $\dot{\gamma}$. If $\omega(0,k_y)>0$ for any wave vector $k_y$, this indicates that the uniform fluid is unstable and will form lanes under shear. Furthermore, the wavenumber for which $\omega(0,k_y)$ is maximal, $k_y=k_y^{max}$, is the fastest growing wavenumber and so this is the wavenumber determining the wavelength of the density oscillations due to the laning. When the amplitude of the density oscillations is large, the wavelength is not exactly $2\pi/k_y^{max}$, since the nonlinear contributions to Eq.\ (\ref{DDFT}) in general slightly shift the wavelength from the value predicted by the linear stability analysis, but generally this shift is small.

\begin{figure}[t!]
\includegraphics[scale=.47]{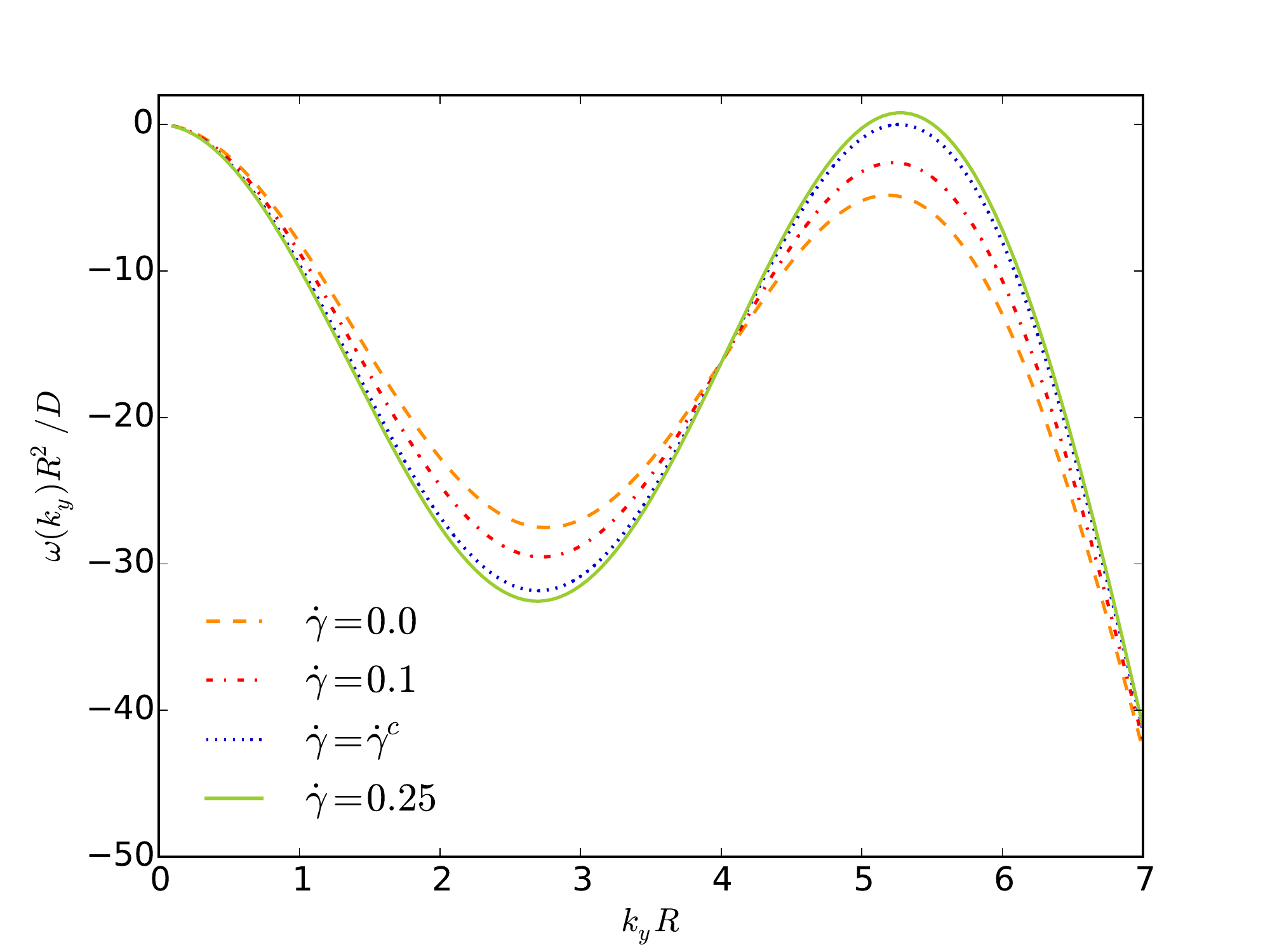}
\caption{Dispersion relation $\omega(0,k_y)$ for various different shear rates $\dot{\gamma}$ for a bulk 2D GEM-8 fluid with density $\rho_b R^2=2.8$ and temperature $k_BT/\epsilon=1$. As $\dot{\gamma}$ is increased, the peak in $\omega(0,k_y)$ moves to larger values, till eventually at the critical shear rate $\dot{\gamma}^c=0.2149$ the uniform liquid becomes linearly unstable, indicating the location of the laning transition.}\label{fig_dispersion}
\end{figure}

\begin{figure}[t!]
\includegraphics[scale=.47]{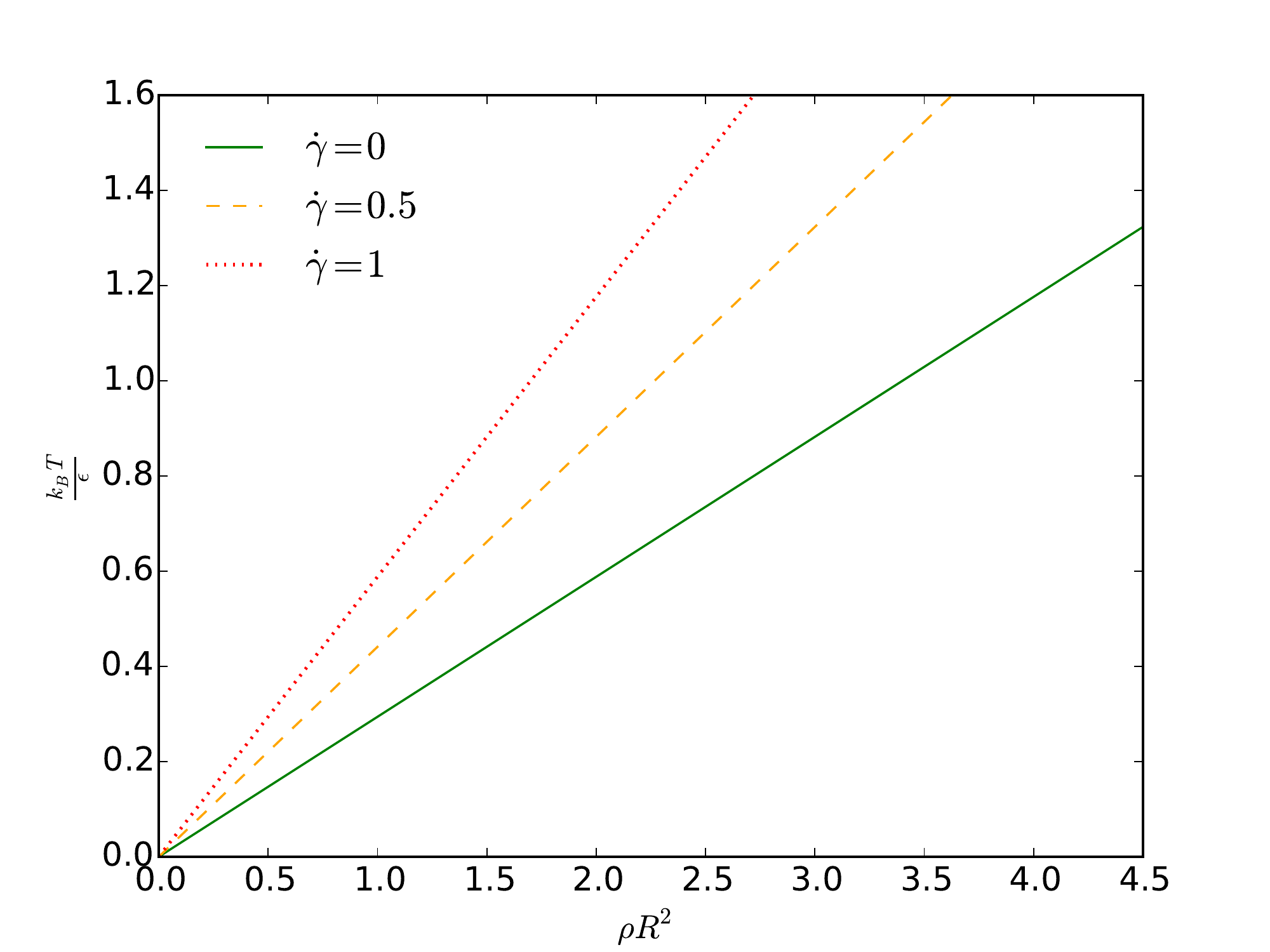}\caption{Linear stability threshold lines (spinodals) for different values of the shear rate $\dot{\gamma}$. For a given $\dot{\gamma}$, states to the left of the line are linearly stable. In contrast, when the system is prepared such that the state point is on the right hand side of the line, the system is unstable.}\label{stability_lines}
\end{figure}

In Fig.~\ref{fig_dispersion} we show a typical example of how the dispersion relation varies as $\dot{\gamma}$ is increased. The results are for a 2D GEM-8 fluid with bulk density $\rho_bR^2=2.8$ and temperature $k_BT/\epsilon=1$. For $\dot{\gamma}=0$ the uniform density liquid is the equilibrium stable state, which can also be seen from the fact that the dispersion relation is strictly negative, i.e.\ $\omega(0,k_y)<0$ for all $k_y$. Increasing $\dot{\gamma}$ we see the peak in $\omega(0,k_y)$ at $k_y=k_y^{max}\approx5.3R^{-1}$ grows and eventually $\omega(0,k_y^{max})=0$ when $\dot{\gamma}=\dot{\gamma}_c=0.2149$, the critical shear rate. For the GEM-8 model we find that the value of $k_y^{max}$ at the laning transition threshold is independent of the density and the temperature $k_BT/\epsilon$. Thus, it is straightforward to calculate the critical shear rate for a given state point $(\rho_b,k_BT/\epsilon)$. In Fig.~\ref{stability_lines} we display the location in the phase diagram of the linear stability threshold calculated for three different values of the shear rate. This is obtained by solving the pair of simultaneous equations
\begin{equation}
\omega(0,k_y)=\frac{d\omega(0,k_y)}{dk_y}=0
\end{equation}
for the critical shear rate $\dot{\gamma}_c$ and the critical wave number $k_y=k_y^{max}$. Note that the linear stability threshold for $\dot{\gamma}=0$ is simply the spinodal associated with the freezing transition, that is discussed in Refs.\ \citep{likos2001effective, mladek2006formation, moreno, andy_walters_thiele_knobloch, andy_alexandr}. The laning transition thresholds are straight lines, passing through the origin, as explained for the case $\dot{\gamma}=0$ in \citep{archer2016generation}. This is a consequence of using the simple mean-field approximation for the excess free energy (\ref{meanfield}), which is accurate at large values of $k_BT/\epsilon$, but when this quantity is small, this approximation becomes unreliable.

\subsection{Density profiles for $\dot{\gamma}$ close to $\dot{\gamma}_c$}\label{subsec:5.C}

\begin{figure}[t!]
\includegraphics[scale=.47]{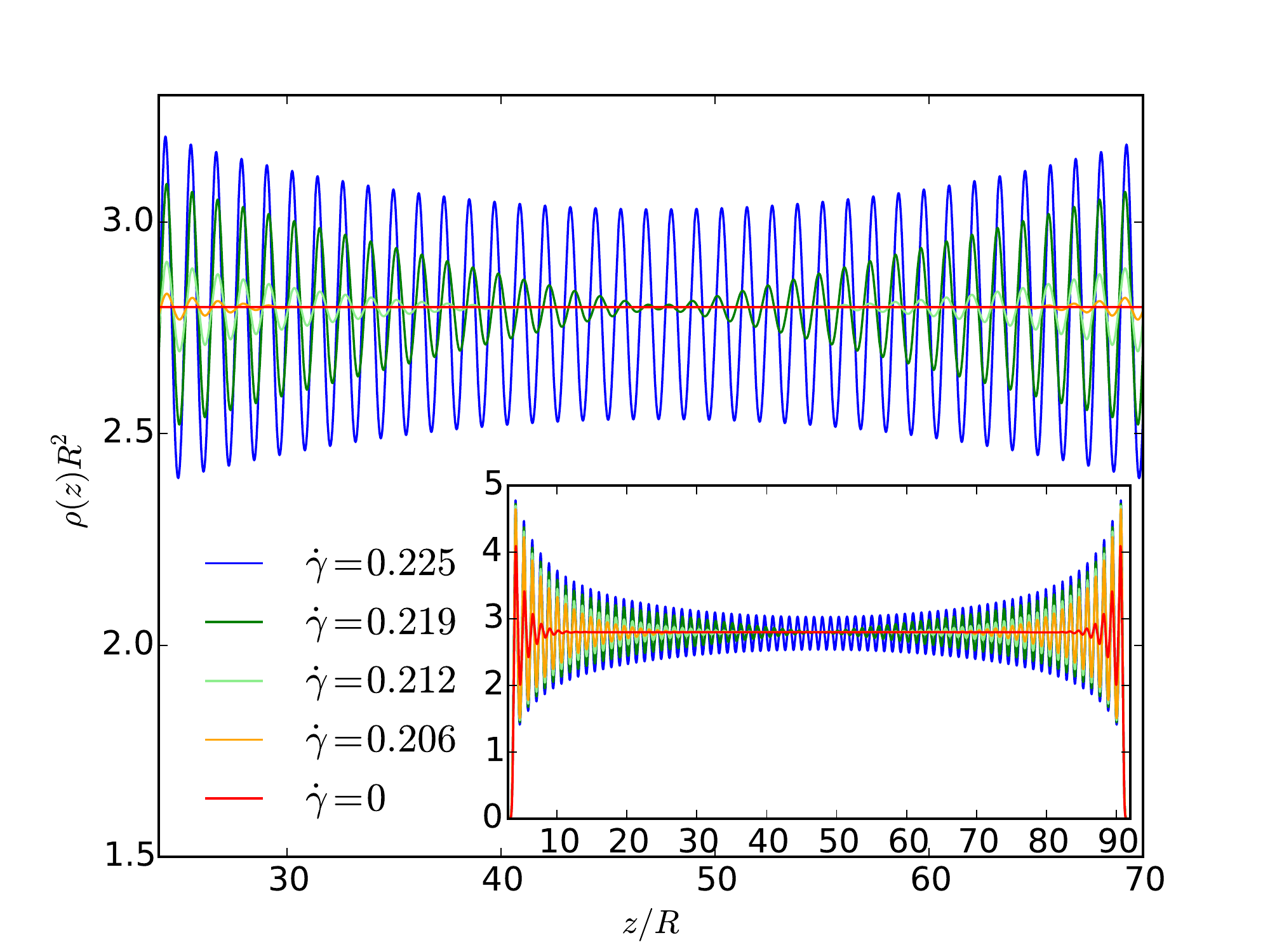}\caption{Fluid density profiles for varying shear rates $\dot{\gamma}$. The bulk fluid average density $\bar{\rho}R^2=2.8$. The critical value derived by the stability analysis is $\dot{\gamma}_c=0.2149$. We see that for shear rates $\dot{\gamma}>\dot{\gamma}_c$, the uniform density is unstable, and the fluid density distribution becomes oscillatory, corresponding to the particles forming distinct layers.}\label{densities}
\end{figure}

In Figs.\ \ref{2D_eq} and \ref{2D_1D} we display density profiles for the 2D GEM-8 fluid for values of $\dot{\gamma}<\dot{\gamma}_c$. We now also display in Fig.\ \ref{densities} density profiles for $\dot{\gamma}$ approaching close to $\dot{\gamma}_c$ and for $\dot{\gamma}>\dot{\gamma}_c$. These are obtained by assuming that the density profile only varies in the $y$-direction, i.e.\ that $\rho(\textbf{r})=\rho(y)$, and then solving the advected-DDFT (\ref{DDFT}) as described above. We see that for $\dot{\gamma}>\dot{\gamma}_c$ the oscillations in the density profile have a finite amplitude at all distances from the confining walls, i.e.\ such states are beyond the laning transition point. The value of $\dot{\gamma}$ where the laning transition occurs in these DDFT calculations is in good agreement with were the transition is predicted to be using the linear stability analysis. There is a small difference that is due to finite size effects, since this is a confined system. The small difference becomes even smaller as the system size is reduced.

As $\dot{\gamma}$ is increased to approach $\dot{\gamma}_c$ from below, we observe in Fig.\ \ref{densities} the presence of density oscillations further and further from the walls. For the static situation, i.e.\ $\dot{\gamma}=0$, the density oscillations are only present close to the confining walls and the amplitude of the oscillations decays relatively fast as one moves away from the walls into the bulk fluid. However, as $\dot{\gamma}\to\dot{\gamma}_c^-$ the decay in the amplitude of the oscillations becomes slower and slower. In fact, the amplitude decay length of the density oscillations diverges as $\dot{\gamma}\to\dot{\gamma}_c^-$. At the critical shear rate for the finite size system in Fig.\ \ref{densities} the density oscillations due to the two confining walls merge together at the mid point of the system and for $\dot{\gamma}>\dot{\gamma}_c$, the particles are ordered in layers right across the system. As $\dot{\gamma}$ is increased above $\dot{\gamma}_c$, the amplitude of the density oscillations in the middle of the system increases continuously from zero, the value for $\dot{\gamma}=\dot{\gamma}_c$.

\section{Discussion and conclusions}
\label{sec:6}

In this paper we have developed a general approach to determine the onset of the laning instability for sheared soft matter suspensions. We have obtained a general expression for the dispersion relation $\omega(0,k_y)$, given in Eq.\ (\ref{dispersion_relation_final}), from which one can find the locus of the laning transition onset as the linear instability line where the maximum of $\omega(0,k_y)=0$. The theoretical framework in which this result is derived is a DDFT incorporating the non-affine particle motion. This stability analysis can be applied to any system where the non-affine contribution to the velocity field, here called $\textbf{v}^{fl}(\textbf{r})$, has the form of Eq.\ (\ref{non-affine}). Note that the present approach is not in any way specific to the soft core GEM-8 model fluid for which results were presented here. We have also applied the present linear stability analysis together with the shear kernels used in the studies of laning in hard-sphere fluids in Refs.\ \citep{brader_kruger_2011, scacchi_kruger_brader_2016}. We do not present any results here, but we do find that the present linear stability analysis predicts the onset of the laning instability to be precisely where it was found in the DDFT calculations.

Whilst the present approach is rather powerful in that it provides a straightforward method to determine where laning occurs, the weak point of the theory is that currently very little is known about the form of the flow kernel $\boldsymbol{\kappa}(\textbf{r}-\textbf{r}')$ in Eq.\ (\ref{non-affine}). Here we use the approximation in Eq.\ (\ref{kernel_grad_phi}) that we believe to be qualitatively correct. Also, Refs.\ \citep{brader_kruger_2011, scacchi_kruger_brader_2016} give good insight into what this quantity is for hard-sphere fluids, but it is clear that much more work is required. We are reminded of the stage in the historical development of liquid state theory where it was realised that the pair direct correlation function $c^{(2)}(r)$ was a key quantity for calculating the equilibrium fluid pair correlations and it was worth working to develop good accurate approximations to it~\citep{hansen_mcdonald}. We are now at a similar stage in the development of the theory for sheared nonequilibrium fluids. We believe the flow kernel is quantity worthy of further study in order to develop more understanding and good approximations for this quantity.

For the state points and shear strengths $\dot{\gamma}$ considered here for the GEM-8 system, we found the fluid density does not vary along the flow direction, parallel to the confining walls [c.f\ Fig.\ \ref{2D_eq}]. However, this is not always the case; symmetry breaking along flow direction can occur. This can be understood by considering that since the laning instability corresponds to a shift of the freezing spinodal to lower densities by the shearing, we should expect that for liquid state points that are near to coexistence with the crystal state that shearing might in fact lead to frozen-like states exhibiting peaks in the density profile. Such shear induced freezing has indeed been seen for Poiseuille flow and will be presented elsewhere~\citep{shear_crystal}.

To conclude, we make two connections to other fields of research: the first is to note that similar laned states are also observed in driven granular media \citep{aranson2006patterns}. Therefore there may be scope to apply the present approach to these systems and we expect that investigating the similarities and differences between granular media and driven colloidal fluids may well shed light on the nature of the flow kernel. The second connection we wish to make is to the mathematics of dynamical systems and bifurcation theory: In the terminology of these subjects the laning transition is a supercritical bifurcation to a periodic state from the flat uniform density state. However, the crystal state bifurcates subcritically from the uniform liquid state at the spinodal, which indicates the possibility of other bifurcations. Combining symmetry considerations with bifurcation theory will surely be useful tools to further understand the laning transition, especially when the fluid is sheared in a way that is more complex than that considered here.

\section{Acknowledgments}
AS is supported by the Swiss National Science Foundation under the grant number 200021-153657/2 and  AJA by the EPSRC under the grant EP/P015689/1.

\begin{appendix}
\section{Reduced flow kernel for the GEM-8 system}
\label{appendix}

We consider first the 3D GEM-8 fluid. Changing the coordinate system from Cartesian to cylindrical polar coordinates, Eq.\ (\ref{non-affine}) together with Eq.\ (\ref{kernel_grad_phi}), with $f=1$, becomes:
\begin{equation}
\begin{split}
\textbf{v}^{fl}(\eta,y,\theta)&
=-\dot{\gamma}\beta R^2\int\int\int \eta' \rho(\eta-\eta',y-y',\theta-\theta')\\&\times \nabla^{cyl} \phi(\eta',y',\theta') d\eta' dy' d\theta'\\&
=-\dot{\gamma}\beta R^2\int\int\int \eta' \rho(\eta-\eta',y-y',\theta-\theta') \\&\times \left[\hat{\boldsymbol{\eta}}\partial_{\eta'}+\hat{\boldsymbol{\theta}}\frac{1}{\eta'}\partial_{\theta'}+\hat{\textbf{y}}\partial_{y'}\right]\phi(\eta',y',\theta')d\eta' dy' d\theta',
\end{split}
\end{equation}
where $\eta^2=x^2+z^2$. The GEM-8 pair potential can be written as $\phi(\eta,y)=\epsilon e^{-\frac{(\eta^2+y^2)^4}{R^8}}$. We now assume the density only varies along the $y$-axis, allowing us to write
\begin{equation}
\begin{split}
\textbf{v}^{fl}(\eta,y)
&=-2\pi \dot{\gamma}\beta R^2 \\ &\times \int\int \eta' \rho(y-y') \left[\hat{\boldsymbol{\eta}}\partial_{\eta'}+\hat{\textbf{y}}
\partial_{y'}\right]\phi(\eta',y')d\eta' dy'.
\end{split}
\end{equation}
Note that the derivative of the pair potential:
\begin{equation}\label{partial_derivative}
\partial_{i}^{j}\phi(i,j)=-i 8(i^2+j^2)^3 \frac{\phi(i,j)}{R^8}\>\>\>,\>\> i,j=\eta,y\>\>,\>\>\eta\neq y,
\end{equation}
where $j$ has to be considered as a constant term. The $y$-component of the kernel function can then be written as:
\begin{equation}
\begin{split}
v^{fl}_y(y)&=\frac{16\pi\dot{\gamma}\beta}{R^6}\int_{-\infty}^{\infty} \rho(y-y') y' \\&\times \int_{0}^{\infty}\eta'(y'^2+\eta'^2)^3\phi(\eta',y')d\eta' dy' \\&
=\dot{\gamma}\int_{-\infty}^{\infty} \rho(y-y') \mathcal{L}_{3D}(y') dy'.
\end{split}
\end{equation}
The form of $\mathcal{L}_{3D}$ is displayed in Fig.~(\ref{kernels}).

In a manner entirely anlogous to that just described for the 3D GEM-8 fluid, for the 2D GEM-8 fluid we can rewrite Eq.\ (\ref{non-affine}) together with Eq.\ (\ref{kernel_grad_phi}) and with $f=1$ as:
\begin{equation}
\begin{split}
\textbf{v}^{fl}(x,y)&=-\dot{\gamma}\beta R^2\int\int \rho(y-y')\nabla\phi(x,y) dx'dy'\\&
=-\dot{\gamma}\beta R^2 \\ &\times\int\int \rho(y-y')\left[\hat{\textbf{x}}\partial_{x'}+\hat{\textbf{y}}\partial_{y'}
\right]\phi(x',y')dx'dy'.
\end{split}
\end{equation}
We focus on the $y$ component of the non-affine term. Recalling the result in equation (\ref{partial_derivative}), for $i,j=x,y$, where $x\neq y$, we can write
\begin{equation}
\begin{split}
v_y^{fl}(y)&=\frac{8\dot{\gamma}\beta}{R^6}\int_{-\infty}^{\infty} \rho(y-y')y'\\& \times \int_{-\infty}^{\infty}(x'^2+y'^2)^3\phi(x',y')dx'dy'\\&
=\dot{\gamma}\int_{-\infty}^{\infty} \rho(y-y')\mathcal{L}_{2D}(y')dy'.
\end{split}
\end{equation} 
The form of $\mathcal{L}_{2D}$ is displayed in Fig.~(\ref{kernels}).

\end{appendix}

\bibliography{source}		
\end{document}